\documentclass[graphicx]{elsart}

\usepackage{graphicx}
\begin{document}
\begin{frontmatter}
\title{Entrance channels and alpha decay half-lives of the heaviest elements}

\author{G. Royer, K. Zbiri, C. Bonilla}\footnote{E-mail:royer@subatech.in2p3.fr} 
\address{Laboratoire Subatech, UMR: IN2P3/CNRS-Universit\'e-Ecole des Mines, \\
 4 rue A. Kastler, 44307 Nantes Cedex 03, France}

\begin{abstract}
The barriers standing against the
formation of superheavy elements and their consecutive $\alpha$ decay 
 have been determined in the quasimolecular shape path 
within a Generalized Liquid Drop Model including the proximity effects 
between nucleons in a neck, the mass and charge asymmetry, a precise nuclear radius and
the shell effects given by the Droplet Model.
For moderately asymmetric reactions double-hump potential barriers stand and 
fast fission of compact shapes in the outer well is possible. Very asymmetric reactions lead to one 
hump barriers which can be passed only with a high energy relatively to 
 the superheavy element energy. Then, only the emission of several neutrons or an $\alpha$ particle 
can allow to reach an eventual ground state.
For almost symmetric heavy-ion reactions, there is no more external well and the inner barrier is higher
than the outer one.
Predictions for partial $\alpha$ decay half-lives are given.

\noindent {\small {\em PACS:} 25.70.Jj; 21.60.Ev; 23.60.+e }

\end{abstract}
\end{frontmatter}

\newpage
  
\section{Introduction}
 Using heavy-ion reactions of mean asymmetry  (Ni on Bi and Zn on Pb \cite{hof95,hof96}) and 
 more recently highly asymmetric reactions (Ca on U, Pu and Cm \cite{og99,og00,og00b})
 new very heavy elements have been synthetized. In these very long and difficult experiments the
 number of detected events ($\alpha$ decay chains) is very low and the interpretation of the
 primary experimental data is discussed \cite{arm00,hof00,lov02}. Actually, it is assumed 
that the lower limit for the fission barrier heights of these heaviest elements is around 
6 MeV \cite{itk02}. To form other superheavy nuclei, many other reactions covering all the 
 asymmetry range are planned or
are proposed using other isotopes, inverse kinematics and radioactive nuclear
beams (and targets) \cite{gup01,pet01,kum03}. 

 The purpose of the present work is, firstly, to study  
the potential barriers standing in the quasimolecular shape path 
for these new suggested entrance channels 
 and to look at their dependence on the next proton magic number (114, 120 or 126). 
The potential barriers relative to already realized 
experiments have been determined in a precedent paper \cite{rogh02}.  
The second objective is to determine the partial $\alpha$ decay half-lives of these superheavy elements
from the Q values given by a recent version of the Thomas-Fermi model \cite{ms96}. 

 The Generalized Liquiq Drop Model which has allowed to describe most of the fusion \cite{rr85}  
, fission \cite{rr84,royzb02}, light nucleus \cite{rm01} and $\alpha$ emission \cite{roy00} data has been used 
 once again. The shell effects have been derived from the Droplet Model formulae \cite{mye77}. 

\section{Quasimolecular shapes} 
To describe the entrance channel leading from two spheres to one sphere and the alpha emission
a shape sequence leading rapidly to the formation of a deep neck while keeping almost spherical 
ends has been retained (see Fig. 1). Mathematically, these quasimolecular shapes correspond to 
two joined elliptic
lemniscatoids \cite{rr85} assuming volume conservation. Analytic formulae are available for the main 
shape-dependent functions : volume, surface, distance between mass centres and moment of inertia.
Similar compact shapes have been also adopted to determine the energy of superheavy nuclei \cite{gpg97}.  
For the exit channel these shapes implicitly suppose that the $\alpha$ particle is rather formed at the 
surface of the parent nucleus.   
  
\section{Generalized Liquid Drop Model}
The macroscopic energy is determined within the Generalized Liquid Drop Model previously defined and used to
describe the fusion and fission processes as well as the 
light nucleus and $\alpha$ emissions \cite{rr85,rr84,royzb02,rm01,roy00}.
\begin{equation}
E=E_{Vol}+E_{Surf}+E_{Coulomb}+E_{Prox}.
\end{equation}  
When the colliding nuclei or the spherical fragments are separated:
\begin{equation}
E_{Vol}=-15.494\left \lbrack (1-1.8I_1^2)A_1+(1-1.8I_2^2)A_2\right 
\rbrack \  MeV,
\end{equation}
\begin{equation}
E_{Surf}=17.9439\left \lbrack(1-2.6I_1^2)A_1^{2/3}+(1-2.6I_2^2)A_2^{2/3}
\right \rbrack \  MeV,
\end{equation}
\begin{equation}
E_{Coulomb}=0.6e^2Z_1^2/R_1+0.6e^2Z_2^2/R_2+e^2Z_1Z_2/r,
\end{equation}  
where $A_i$, $Z_i$, $R_i$ and $I_i$ are the masses, charges, radii and relative 
neutron excesses of the two nuclei. r is the distance between the mass centres.\newline  
The radii $R_{i}$ are given by:
\begin{equation}
R_i=(1.28A_i^{1/3}-0.76+0.8A_i^{-1/3}) \  fm.
\end{equation}
This formula allows to follow the experimentally observed increase
of the ratio $r_i=R_i/A_i^{1/3}$ with the mass; for example, $r_0=1.13fm$
for $^{48}Ca$ and $r_0=1.18fm$ for $^{248}Cm$.\newline       
For one-body shapes, the volume, surface and Coulomb energies are defined as
\begin{equation}
E_{Vol}=-15.494(1-1.8I^2)A \ MeV,
\end{equation}
\begin{equation}
E_{Surf}=17.9439(1-2.6I^2)A^{2/3}(S/4\pi R_0^2) \ MeV,
\end{equation}
\begin{equation}
E_{Coulomb}=0.6e^2(Z^2/R_0) \times 0.5\int (V(\theta)/V_0)(R(\theta)/R_0)^3
\sin \theta d \theta.
\end{equation}  
$I$ is the relative neutron excess and $S$ the surface of the  
one-body deformed nucleus. $V(\theta )$ is the electrostatic potential
 at the surface and $V_0$ the surface potential of the sphere. The 
radius of the compound nuclear system has been calculated from
the radii of the two fragments assuming volume conservation.    

The surface energy results from the effects of the surface
tension forces in a half space. When there are 
nucleons in regard in a neck or a gap between colliding nuclei or separated fragments
an additional term called proximity energy must be added 
 to take into account the effects of the nuclear forces between the close surfaces in regard.
 This term is essential to describe smoothly the two-body to one-body transition
and to obtain reasonable fusion barrier heights. It moves the barrier top to
an external position and strongly decreases the pure Coulomb barrier.
\begin{equation}
E_{Prox}(r)=2\gamma \int _{h_{min}} ^{h_{max}} \Phi \left \lbrack D(r,h)/b\right 
\rbrack 2 \pi hdh,
\end{equation}  
where $h$ is the distance varying from the neck radius or zero
to the height of the neck border. $D$ is the distance between the 
surfaces in regard and $b=0.99 fm$ the surface width. 
 $\Phi$ is the proximity function of Feldmeier \cite{fel79}. The surface 
 parameter $\gamma$ is the geometric mean between the surface parameters 
 of the two nuclei or fragments:  
\begin{equation}
\gamma=0.9517\sqrt{(1-2.6I_1^2)(1-2.6I_2^2)} \  MeVfm^{-2}.
\end{equation}  
In this GLDM the surface diffuseness is not taken into account and the 
proximity energy vanishes when there is no neck or gap.
The combination of these GLDM and shape sequence has allowed to reproduce  
the fusion barrier heights and radii, the fission and the $\alpha$ 
and cluster radioactivity data.              

\section{Potential barriers governing the entrance channel}
The macroscopic potential barriers (deformation energy) standing in front of 
some heavy-ion combinations leading to $^{270}110$ and $^{302}120$ compound systems
are displayed in Fig. 2. Some reactions using very exotic nuclei are not still feasible.
Within this GLDM the nuclear proximity energy introduces an inflection in the curve
and the outer barrier top corresponds always 
 to an unstable equilibrium
between the repulsive Coulomb forces and the attractive proximity forces for two separated nuclei.
 For the highest asymmetries the external fusion barrier energy is higher than the spherical system
energy and a potential pocket appears after crossing the barrier. In this description, the nuclear system
can reach a quasi-spherical shape with little excitation energy via a tunneling process. 
This does not prove the stability of the formed system. For the exit channel all the symmetric and asymmetric
fission and $\alpha$ emission barriers must be investigated.
With decreasing asymmetry the outer well progressively disappears as the outer peak.  
It will be very difficult to reach the
eventual ground state of the superheavy element via these almost symmetric reactions.
Predictions from such barriers for four asymmetric and almost symmetric reactions leading to $^{246}Fm$ 
had been compared in a previous paper \cite{rr85} with experimental data on evaporation residues and fast 
fission events and the agreement was very good.   
 
In ordinary fusion studies, it is only the preceding barriers which are taken into account. The 
macroscopic energy does not reproduce very accurately the Q value since the nonuniform distribution
 of single-particle levels leading to the shell correction term and pairing energy is not taken into account
 as also Wigner energy,...
 In Fig. 3, the difference between the theoretical Q value given by the Thomas-Fermi model \cite{ms96} 
and the GLDM one has been empirically added at the
macroscopic energy of the compound nucleus with a linear (in r) attenuation factor vanishing at the contact point. 
 Such a description is valid if the nuclear system has enough time to relax and built its shells 
and pairs during its descent to 
a quasi-spherical shape. Then the barrier against reseparation
is higher and the external minimum disappears for the highest asymmetries.

In the following the shell corrections proposed in the Droplet Model \cite{mye77,rr84} will be used and the formulae
are recalled here.   
\begin{equation}
E_{Shell}=E_{Shell}^{sphere}(1-2\theta^{2})e^{-\theta^{2}}.
\end{equation}
The shell corrections for a spherical nucleus are given by
\begin{equation}
E_{Shell}^{sphere}=5.8\left \lbrack (F(N)+F(Z))/(0.5A)^{2/3}-0.26A^{1/3}\right 
\rbrack \  MeV,
\end{equation}
where, for $M_{i-1}<X<M_i$,
\begin{equation}
F(X)=q_i(X-M_{i-1})-0.6(X^{5/3}-M_{i-1}^{5/3}).
\end{equation}
$M_i$ are the magic numbers and
\begin{equation}
q_i=0.6(M_i^{5/3}-M_{i-1}^{5/3})/(M_i-M_{i-1}).
\end{equation}
Within this algebraic approach it is possible to study the influence of the selected value
for the highest proton magic number. For the two highest neutron magic numbers the values 126 and 184
have been retained.

\begin{equation}
\theta^{2}=({\delta}R)^2/a^2.
\end{equation}
The distortion ${\theta}a$ is the root mean square of the deviation of the nuclear surface from the sphere, 
a quantity which incorporates indiscriminately all types of deformation. The range a 
has been chosen to be 
$0.32r_0$.

Potential barriers for three very asymmetric reactions ('warm' reactions)
$^{48}Ca$+$^{244}Pu$, 
 $^{48}Ca$+$^{248}Cm$ and $^{50}Ti$+$^{248}Cm$  
are shown in Fig. 4 assuming different hypotheses. Due to the high asymmetry and, consequently,
to a low Coulomb repulsion and proximity energy the barrier against reseparation is wide and high. 
There is no double-hump barriers. 
Even for a subbarrier tunneling of 6 or 7 MeV in the entrance channel
and even if the shells and pairs have not enough time to develop
(dashed curve) before reaching the quasispherical shape 
the nuclear system has enough energy to reach this quasispherical compound system. 
The excitation energy relatively to the ground state energy is more than 30 MeV, allowing 
the emission of several neutrons or an $\alpha$ particle.
The different hypotheses on the proton magic number do not change the global 
predictions in the entrance path.

Potential barriers governing the medium asymmetry reactions ('cold' reactions)
$^{64}Ni$+$^{208}Pb$, $^{70}Zn$+$^{208}Pb$, $^{76}Ge$+$^{208}Pb$ 
and $^{82}Se$+$^{208}Pb$ are displayed in Fig. 5. A wide macroscopic potential pocket due 
mainly to higher proximity energy and Coulomb repulsion (relatively to warm reactions) appears 
at large deformations. Whatever the microscopic correction assumptions are double-hump barriers
begin to appear for the Ni on Pb and Zn on Pb reactions. Experimentally the beam energy was respectively
239 and 257 MeV. Consequently, these successfull experiments correspond to subbarrier fusion processes
in our approach and the quasispherical system can be reached by tunneling even if the shells and pairs are
not completely built. For the two last reactions the inner macroscopic hill is higher than
the external barrier. Incomplete fusion and fast fission events in the external
pocket are the main exit channels since the neck between the two nuclei is formed and exchanges of nucleons may occur. 
If the reorganization of the single-particle levels is very rapid then the value of the proton magic 
number begins to play some role. So an open question is whether at large deformations
the nucleon shells can take form to stabilize the nuclear system before 
investigating a peculiar exit channel.
 The pre or post equilibrium nature of the evaporation process
 of the excess neutron is also crucial. 

 The same comments are valid for the two cold reactions shown 
in Fig. 6 leading to nuclei of charge 118 and 119. A value of 120 for the next proton magic number
naturally would slightly increase the possibility to form such superheavy elements.     
For still heavier mass and charge and more symmetric heavy-ion reactions (see Fig. 7) the potential
pocket between the two peaks progressively disappears. A tunneling effect in the outer peak does not
allow to reach a quasispherical state. Only a subbarrier fusion through the inner peak could eventually
lead to a compound nucleus if the microscopic contributions grow rapidly at  the beginning of the fusion 
process and if the charge of the compound system is the next proton magic number.

The symmetric decay barriers for $^{283}112$, $^{292}116$, $^{295}120$ and $^{311}126$ 
are given in Fig. 8. The microscopic contributions play the main role. The height and width of the one hump
fission barrier is governed by the proximity of a proton or neutron magic number. The barrier height still reaches
several MeV and can reach 10 MeV if the neutron and proton shell effects add each other.

The characteristics of the macroscopic potential barriers for the new suggested reactions 
\cite{gup01,pet01,kum03} leading possibly 
to superheavy elements are shown in table 1.
The Q value is extracted from \cite{ms96}. The energy of the external peak relatively to the sphere
energy is positive for $Z_1Z_2$ lower than around 2500. The position of the external barrier corresponds
to the contact point of the colliding nuclei when $Z_1Z_2$ is higher than around 2600. There is no more real 
external macroscopic barrier for $Z_1Z_2$ higher than around 3200. 

In the hope of reaching the most efficient beam energy, formulas giving the
l-dependent fusion barrier height $E_{fus}$ and radius $R_{fus}$ are
proposed.
\begin{eqnarray}
E_{fus,l}(MeV) & = & E_{fus,l=0}  \\
& + &   \frac{l(l+1)}{0.02081(A_1^{5/3}+A_2^{5/3})+
\frac{0.0506A_1A_2(A_1^{1/3}+A_2^{1/3})^2(1.908+\frac{3.94}{Z_1Z_2}
-0.0857ln(Z_1Z_2))^2}{A_1+A_2}}, \nonumber
\end{eqnarray}    
with 
\begin{equation}                 
E_{fus,l=0}(MeV)=-19.38+ 
\frac{2.1388Z_1Z_2+59.427(A_1^{1/3}+A_2^{1/3})-27.07
\ln(\frac{Z_1Z_2}{A_1^{1/3}+A_2^{1/3}})} 
{(A_1^{1/3}+A_2^{1/3})(2.97-0.12\ln(Z_1Z_2))},
\end{equation}
\begin{eqnarray}
R_{fus,l}(fm) & = & (A_1^{1/3}+A_2^{1/3})\left \lbrack 1.532+6.48/Z_1Z_2\right \rbrack \\
& - &0.000507(A_1^{1/3}+A_2^{1/3})^2ln(Z_1Z_2)^2-\frac{52.8l^2}{A_1^2A_2^2} 
-\frac{1.99l^2}{(A_1+A_2)^2}. \nonumber
\end{eqnarray}  
For frontal collisions the fusion radius is still more precisely given by 
\begin{equation}
R_{fus,l=0}(fm)=(A_1^{1/3}+A_2^{1/3})\left \lbrack 1.908-0.0857ln(Z_1Z_2)+\frac{3.94}{Z_1Z_2}
\right \rbrack.
\end{equation}             

\section{Excitation energy}
Assuming a full equilibrated ground state the experimental excitation energy $E^{*}_{exp}$ 
is the sum of the beam energy $E_{cm}$ 
in the mass centre frame and the $Q_{fusion}$ value. For the heaviest already investigated experiments,
 $E^{*}_{exp}$ is compared, in table 2, with
the energy $E^{*}_{bar}$ needed to pass classically the outer fusion barrier peak given by the GLDM.
The difference between these two values is given in the last column. The negative sign corresponds in our approach 
to a tunneling process in the entrance channel. The range of some MeV below or above the barrier and the 
barrier profiles are consistent with the very low fusion cross sections and the ability to reach the compound nucleus.
Let us recall that the fusion barrier height derived from the GLDM is systematically higher than the Bass barrier
height and the fusion radius is smaller \cite{rogh02}. Consequently, for most of the presently used incident energies,
the fusion is above the barrier for the Bass model and a subbarrier fusion for the GLDM.

\section{$\alpha$ decay half-lives}

To determine the potential barriers against $\alpha$ emission, the $\alpha$ decay 
has been viewed \cite{roy00} as a very asymmetric spontaneous fission within the
GLDM. The difference between the experimental $Q_{\alpha}$ value or the value
predicted by the Thomas-Fermi model \cite{ms96} and the value given by the GLDM has also
been added at the sphere energy with a linear attenuation factor vanishing at the rupture
point. Within such an unified fission model the decay constant of the parent nucleus is
simply defined as $\lambda=\nu_0P$. The assault
frequency  $\nu_0$ has been chosen as $10^{20}\ s^{-1}$. The barrier penetrability P has been
calculated within the general form of the action integral. Most of the decay path corresponds
to two-body shapes and the reduced mass approximation is sufficient for the inertia.  
The predicted $\alpha$ decay half-lives agree with the experimental data in the whole mass 
range and also for the known heaviest elements \cite{lov02,roy00}.       
Accurate formulae depending only on the mass and charge of the parent nucleus and $Q_{\alpha}$ 
have been extracted from the data. They are recalled here respectively
 for the even(Z)-even(N), even-odd, odd-even and odd-odd nuclei.

\begin{equation}
log_{10}\left \lbrack T_{1/2}(s)\right \rbrack=-25.31
-1.1629A^{1/6}\sqrt{Z}+\frac{1.5864Z}{\sqrt{Q_{\alpha}}},
\end{equation}  
\begin{equation}
log_{10}\left \lbrack T_{1/2}(s)\right \rbrack=-26.65
-1.0859A^{1/6}\sqrt{Z}+\frac{1.5848Z}{\sqrt{Q_{\alpha}}},
\end{equation}  
\begin{equation}
log_{10}\left \lbrack T_{1/2}(s)\right \rbrack=-25.68
-1.1423A^{1/6}\sqrt{Z}+\frac{1.592Z}{\sqrt{Q_{\alpha}}},
\end{equation}  
\begin{equation}
log_{10}\left \lbrack T_{1/2}(s)\right \rbrack=-29.48
-1.113A^{1/6}\sqrt{Z}+\frac{1.6971Z}{\sqrt{Q_{\alpha}}}.
\end{equation}  

The global predictions for the  $\alpha$ decay half-lives of the heaviest 
elements within the preceding formulas are given in table 3. If the $\alpha$ decay mode 
is the main decay mode, then for most of them their half-lives will vary from microsecond 
to some days. Nevertheless, some nuclei ($^{319,320}126$ and $^{317}124$,
 i.e Z around 124-126 and N around 193-194) have a very low $Q_{\alpha}$ 
and consequently a very high $\alpha$ decay half-life. About the same results are 
obtained within the $Q_{\alpha}$ value given by the finite-range Droplet 
macroscopic model coupled to the folded-Yukawa single-particle 
microscopic model \cite{moll95}. 
The calculation of the half-lives against symmetric and asymmetric fission is
another challenge.
                       
\section{Conclusion} 
The barriers preventing the
formation of superheavy elements and their consecutive $\alpha$ decay 
 have been determined in the quasimolecular shape path 
within a Generalized Liquid Drop Model including the proximity effects 
between nucleons in a neck and the mass and charge asymmetry. The dependence on the asymmetry in 
the entrance channel, on the proton magic number of the compound system and 
on the introduction of the Q value 
has been studied. The shell effects have been determined within the Droplet Model formulae.

For moderately asymmetric reactions double-hump potential barriers stand and  
fast fission of compact shapes in the outer well is the main exit channel. 
Very asymmetric reactions lead to one 
hump barriers which can be passed only with an energy much higher than the ground state
energy of the superheavy element. Then, only emission of several neutrons 
or an $\alpha$ particle can stabilize the nuclear system and allows to reach a ground state.
The formation of superheavy elements via almost symmetric reactions is hardly likely.  

In this very heavy mass range the Droplet Model formulae lead to fission barrier heights of several MeV
 or even 10 MeV for nuclei close to the 
proton and neutron magic numbers. The main decay mode being the $\alpha$ decay, 
the predicted half-life depends on the selected theoretical $Q_\alpha$  value.
Within the $Q_\alpha$ given by the Thomas-Fermi model, partial $\alpha$ decay half-lives are proposed.
They agree generally with the known experimental data.

\newpage
{\bf Figure captions}

Fig. 1. Selected shape sequence to simulate the transition from two-body
to one-body shapes and the alpha decay path. The nuclei are spherical
when they are separated.
 
Fig. 2. Macroscopic potential barriers for different reactions leading to the 
 $^{270}110$ and $^{302}120$ nuclei. r is the distance between the mass
centres. The vertical bar corresponds to the contact point.  

Fig. 3. Same as figure 2 for $^{302}120$ but an adjustment is done
to reproduce the Thomas-Fermi model Q value. 

Fig. 4. Potential barriers for the $^{48}Ca$+$^{244}Pu$, 
 $^{48}Ca$+$^{248}Cm$, $^{50}Ti$+$^{248}Cm$ reactions. 
The dashed curve corresponds to the macroscopic barrier. 
 The barrier given by the dashed and double dotted line incorporates a linear correction 
from the contact point till the sphere to reproduce the Q value.
The full line, 
dotted curve and dashed and dotted curve include the shell effects given 
by the Droplet Model assuming respectively a proton magic number of 114,
120 and 126 and an adjustment to reproduce the Q value. 

Fig. 5. Same as figure 4 but for the reactions $^{64}Ni$+$^{208}Pb$, 
$^{70}Zn$+$^{208}Pb$, $^{76}Ge$+$^{208}Pb$ and $^{82}Se$+$^{208}Pb$.

Fig. 6. Same as figure 4 but for the reactions $^{86}Kr$+$^{208}Pb$
 and $^{87}Rb$+$^{208}Pb$.

Fig. 7. Same as figure 4 but for the reactions $^{88}Sr$+$^{208}Pb$, 
$^{116}Cd$+$^{181}W$ and $^{104}Ru$+$^{208}Pb$.

Fig. 8. Symmetric fission barriers for $^{283}112$, $^{292}116$, $^{295}120$ and $^{311}126$.
The full line, dotted curve and dashed and dotted curve include the shell effects given 
by the Droplet Model assuming respectively a proton magic number of 114,
120 and 126; the neutron magic numbers being 126 and 184. 

{\bf Table captions}    

Table 1. Characteristics of the macroscopic potential barriers for 
the heaviest nuclear systems.  
$R_{min}$, $R_{ext}$, $H_{min}$ and $H_{ext}$ are 
respectively the positions and energies relatively to the sphere of the 
outer minimum and external maximum of the barrier. 
$R_{12}$ is the distance between the mass centres at the contact point.
$H_{ext, fus}$ is the GLDM external fusion barrier height.

Table 2. Comparison between the experimental excitation energy $E^{*}_{exp}$ and the classical excitation energy
$E^{*}_{bar}$ at the top of the GLDM barrier. In the last column the positive or negative sign of the 
difference $E^{*}_{exp}-E^{*}_{bar}$ indicates  a fusion above or below the barrier.

Table 3. Predicted $log_{10}\lbrack T_{1/2}(s)\rbrack$ for the heaviest elements
and nuclear systems versus the charge and mass of the parent nucleus and the $Q_{\alpha}$ value
given by a recent version of the Thomas-Fermi model \cite{ms96}.

\newpage
\centerline{
\includegraphics[width=10cm]{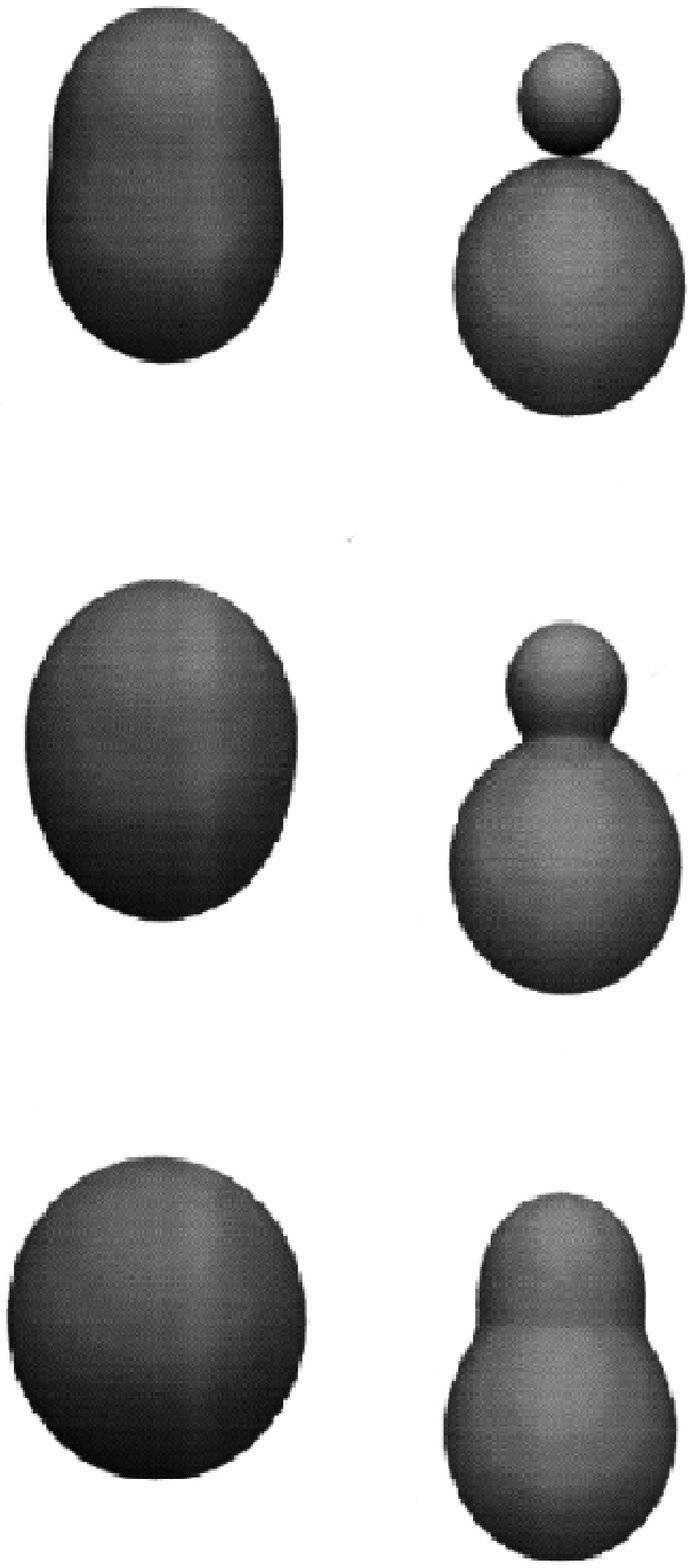}
}
\bigskip

\newpage
\centerline{
\includegraphics[width=18cm]{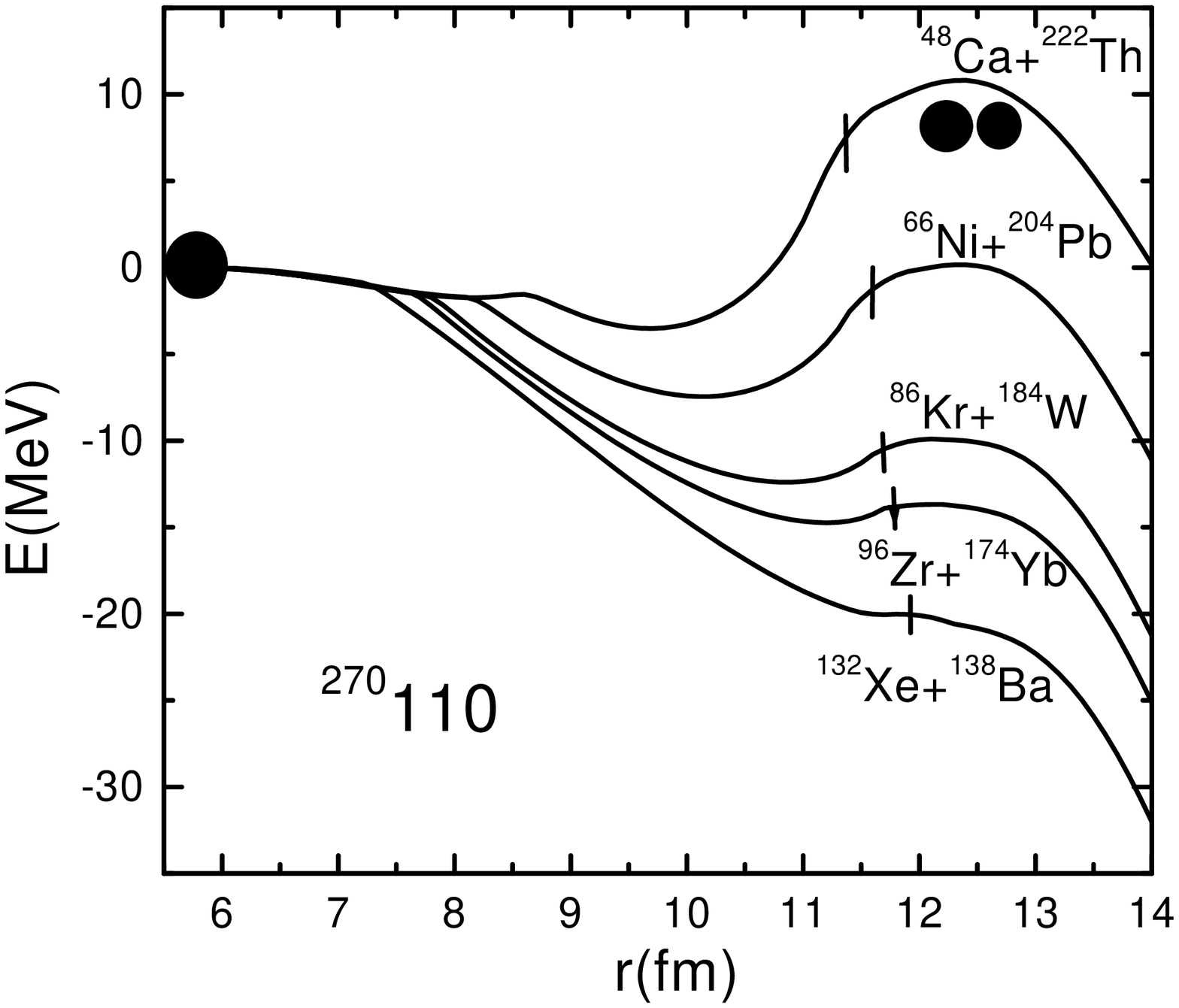}
}
\bigskip

\newpage
\centerline{
\includegraphics[width=20cm]{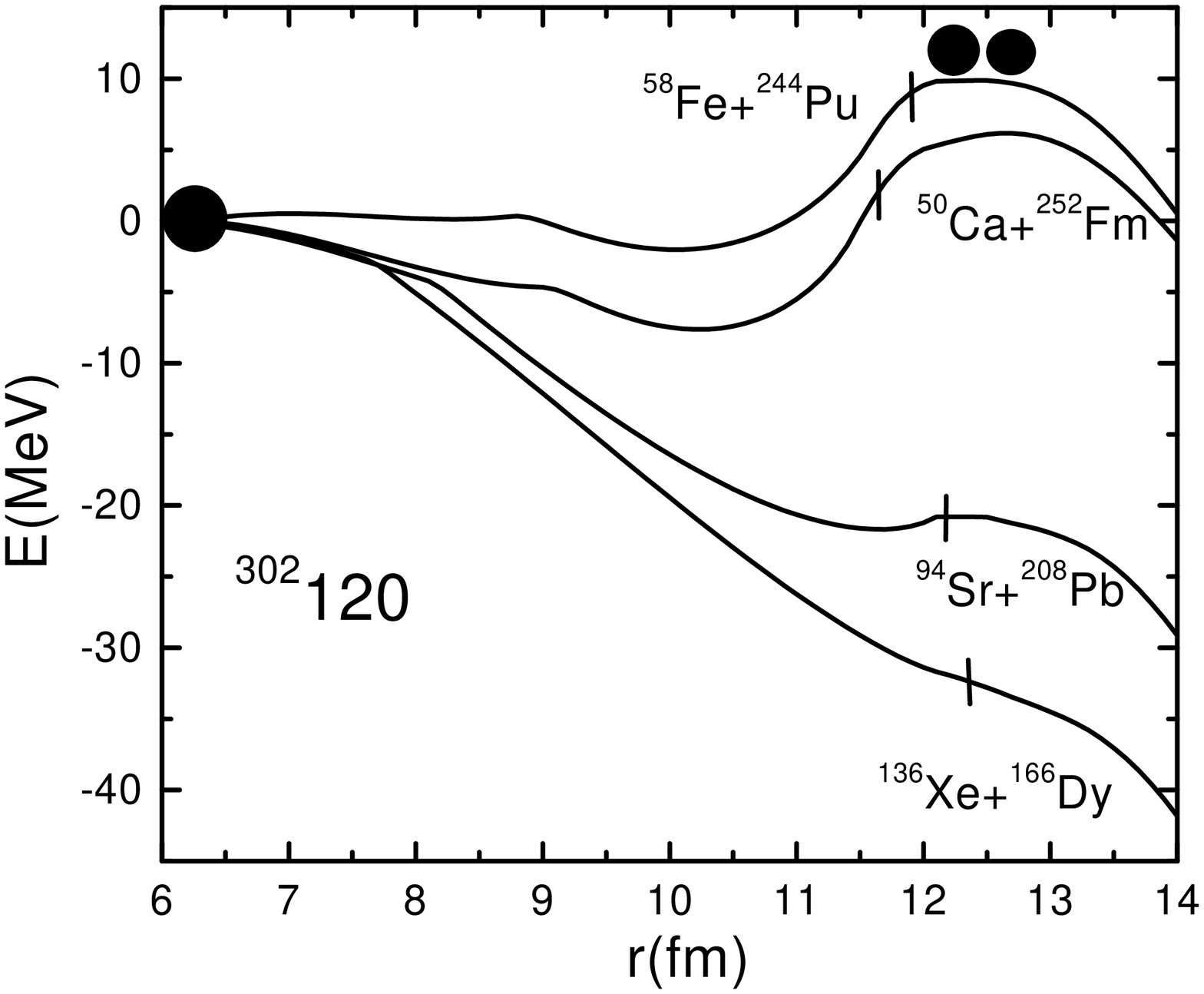}
}
\bigskip

\newpage
\centerline{
\includegraphics[width=20cm]{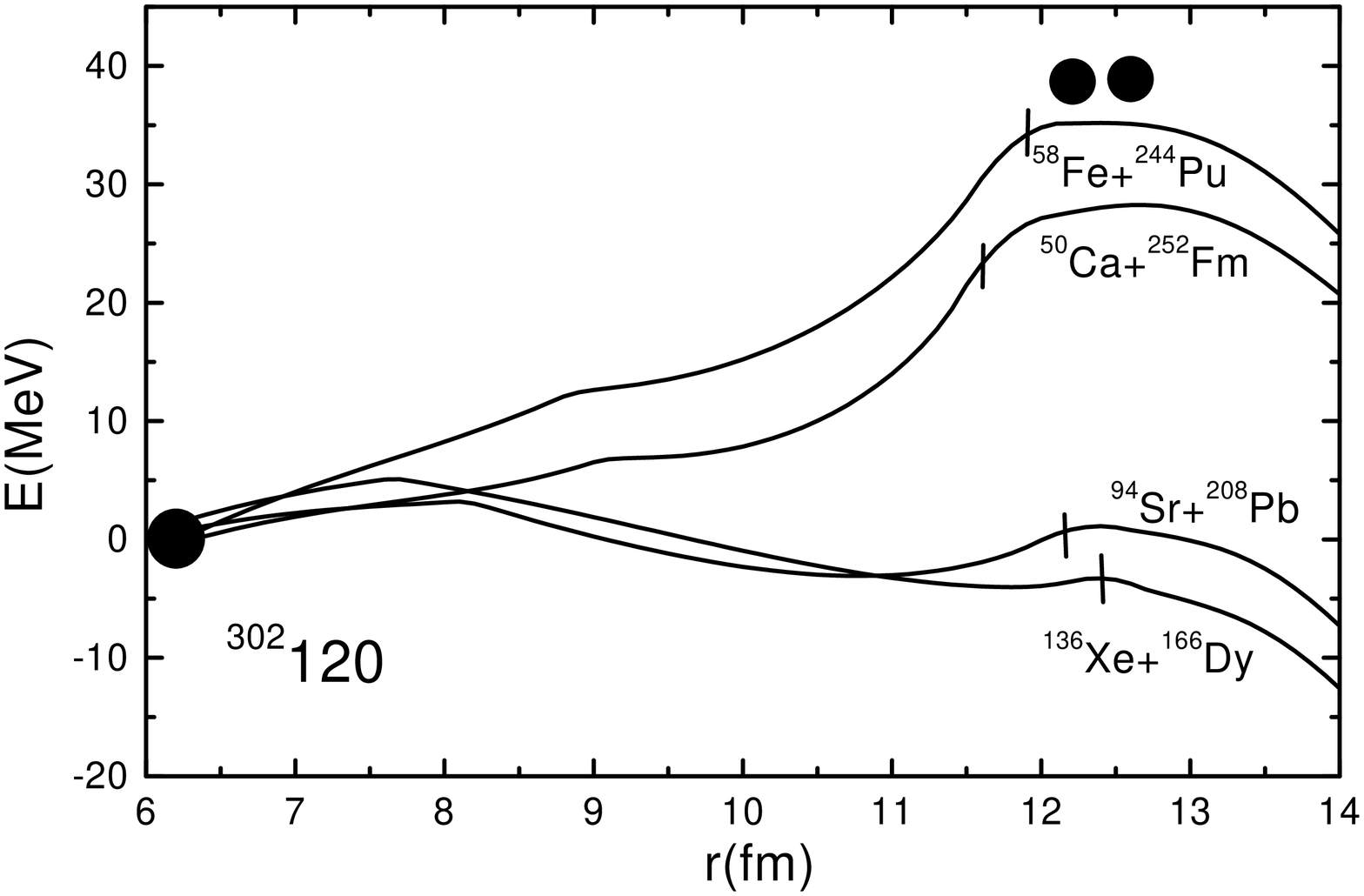}
}
\bigskip

\newpage
\centerline{
\includegraphics[width=20cm]{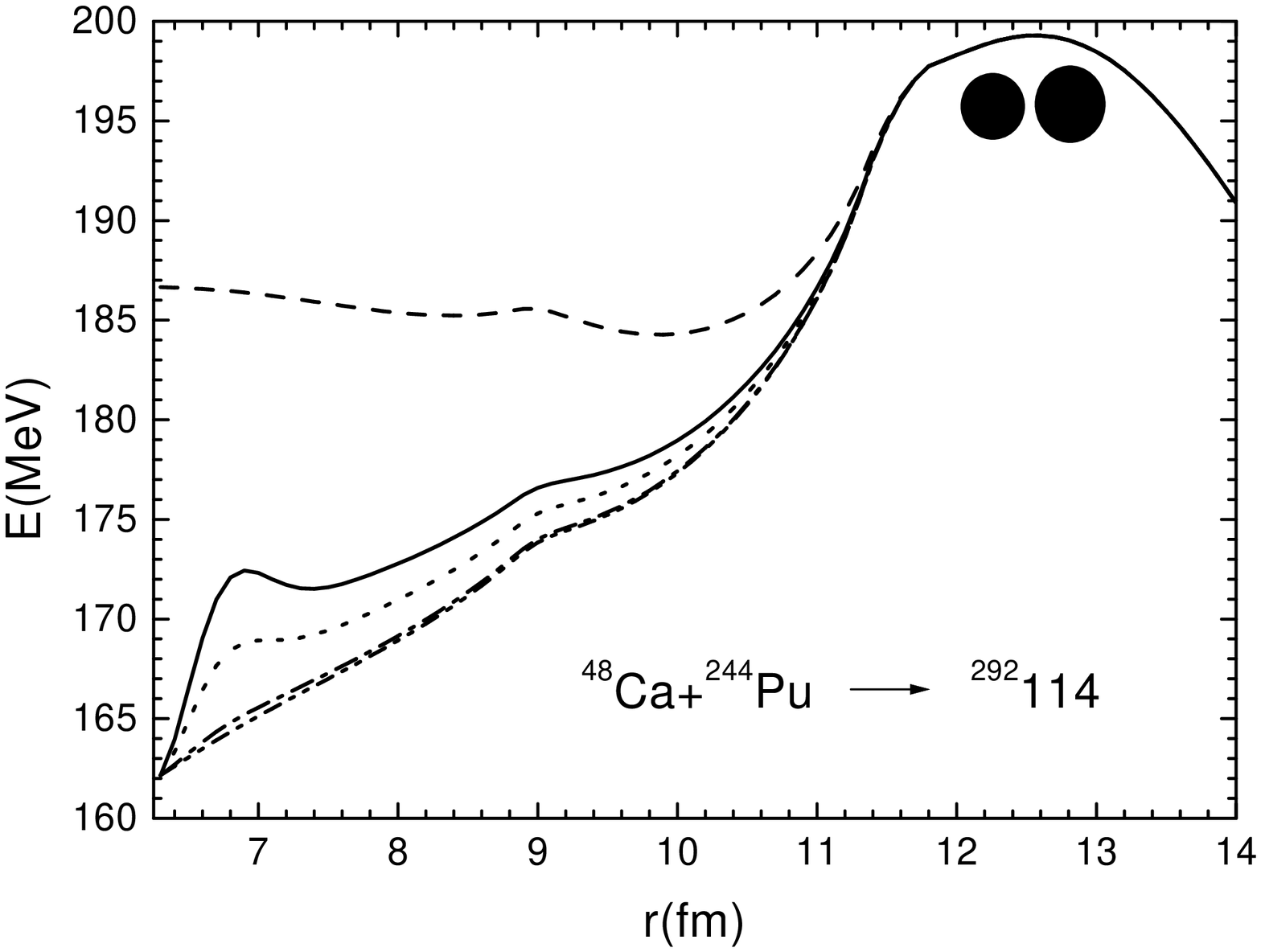}
}
\bigskip

\newpage
\centerline{
\includegraphics[width=20cm]{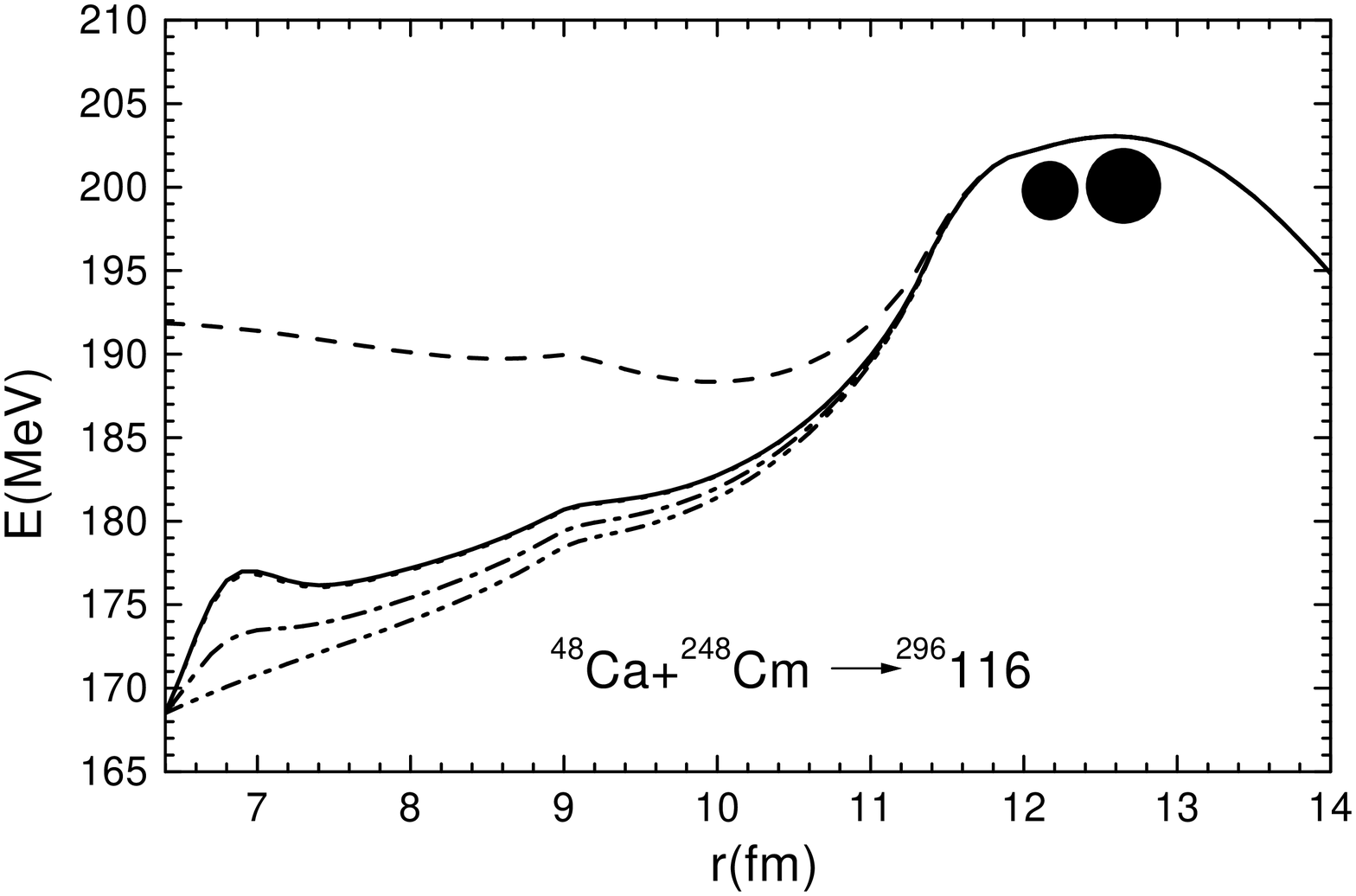}
}
\bigskip

\newpage
\centerline{
\includegraphics[width=20cm]{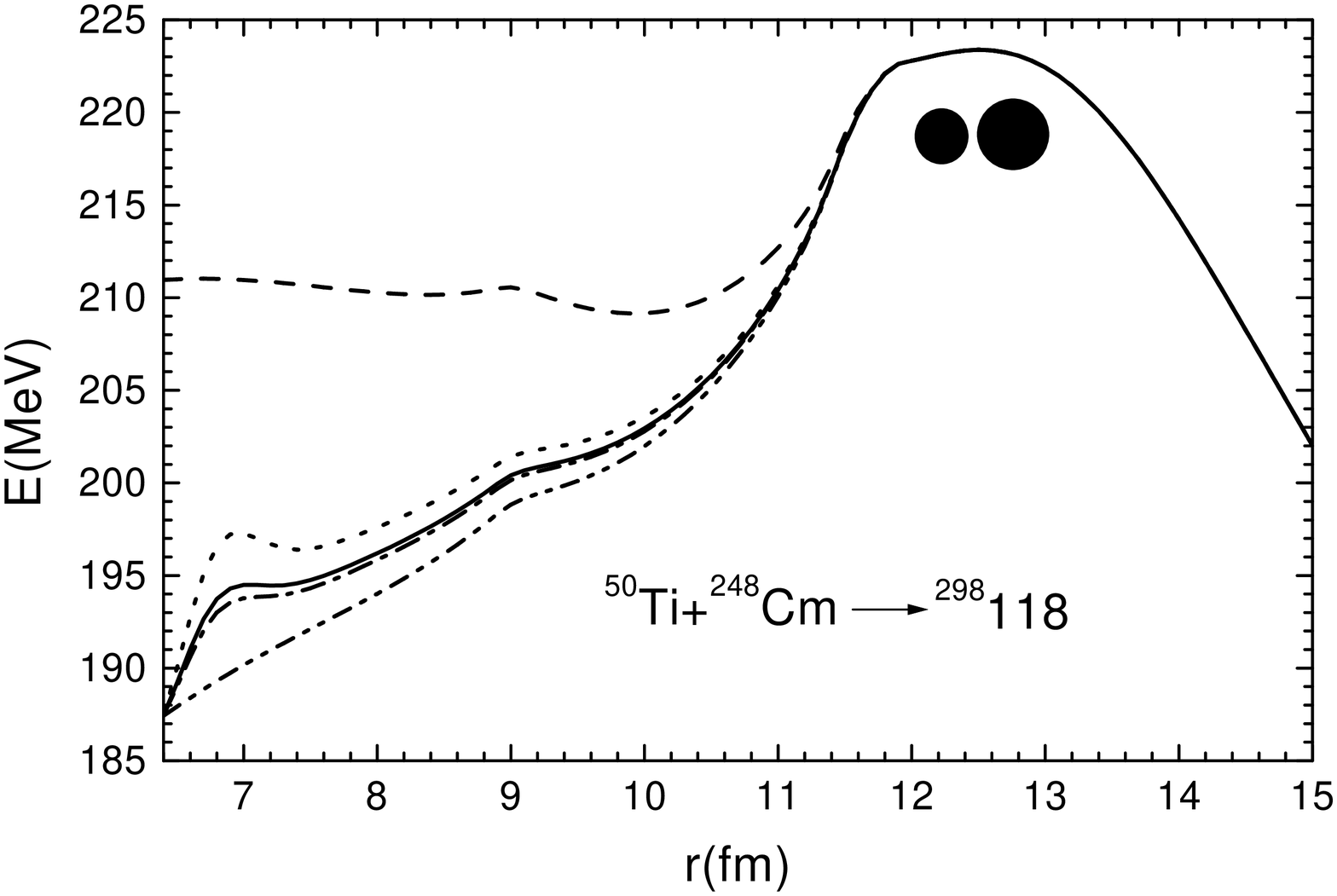}
}
\bigskip

\newpage
\centerline{
\includegraphics[width=20cm]{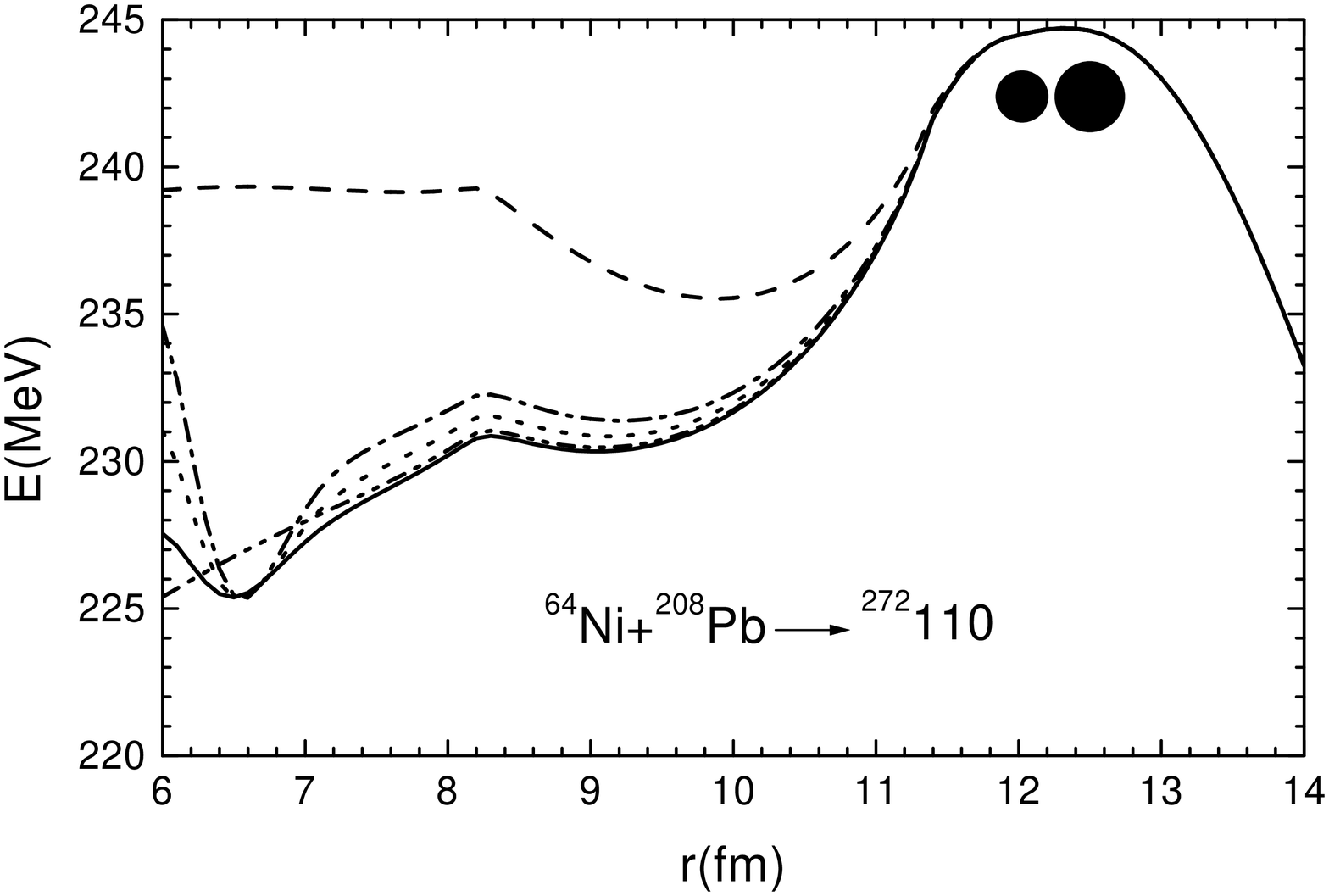}
}
\bigskip

\newpage
\centerline{
\includegraphics[width=20cm]{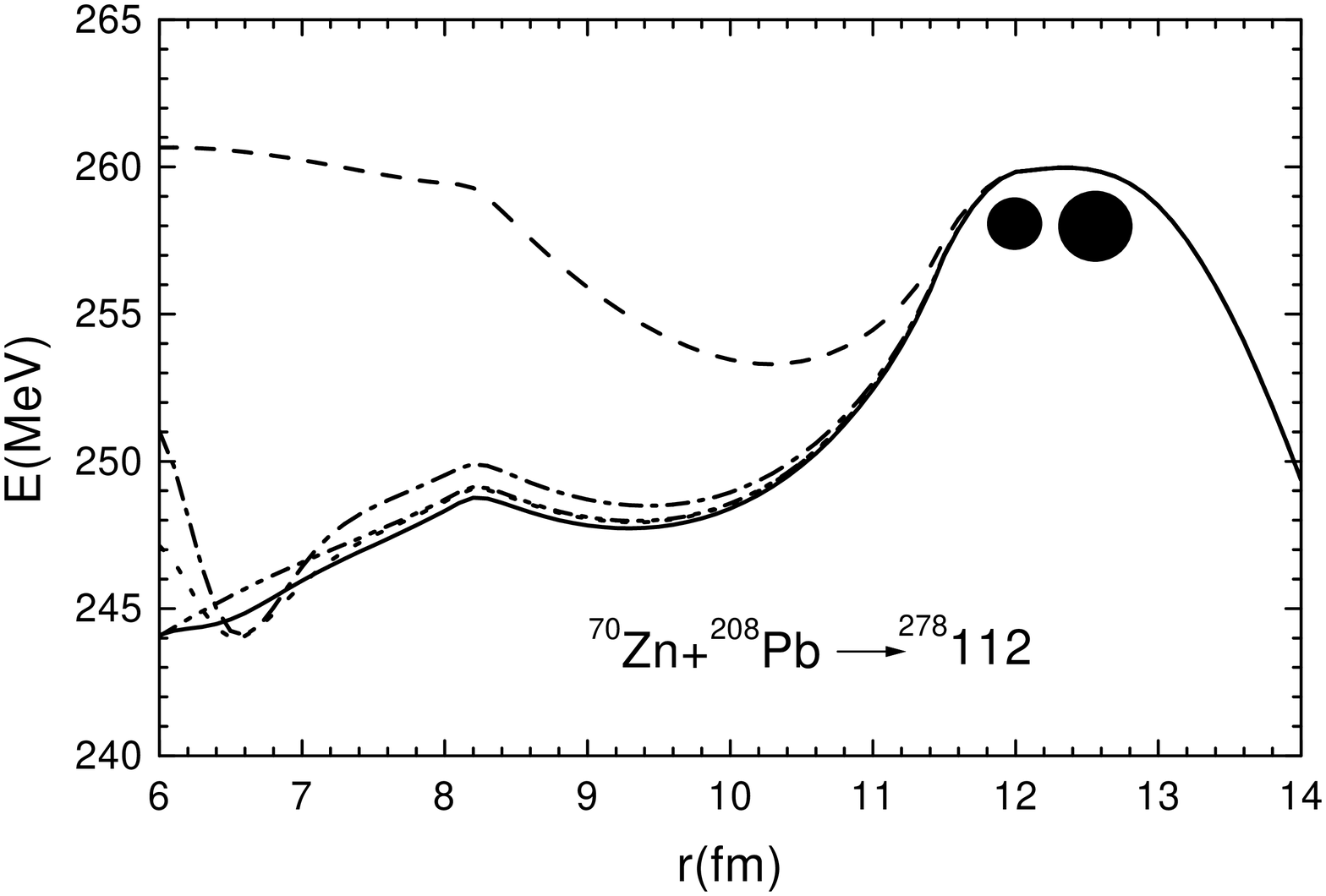}
}
\bigskip

\newpage
\centerline{
\includegraphics[width=20cm]{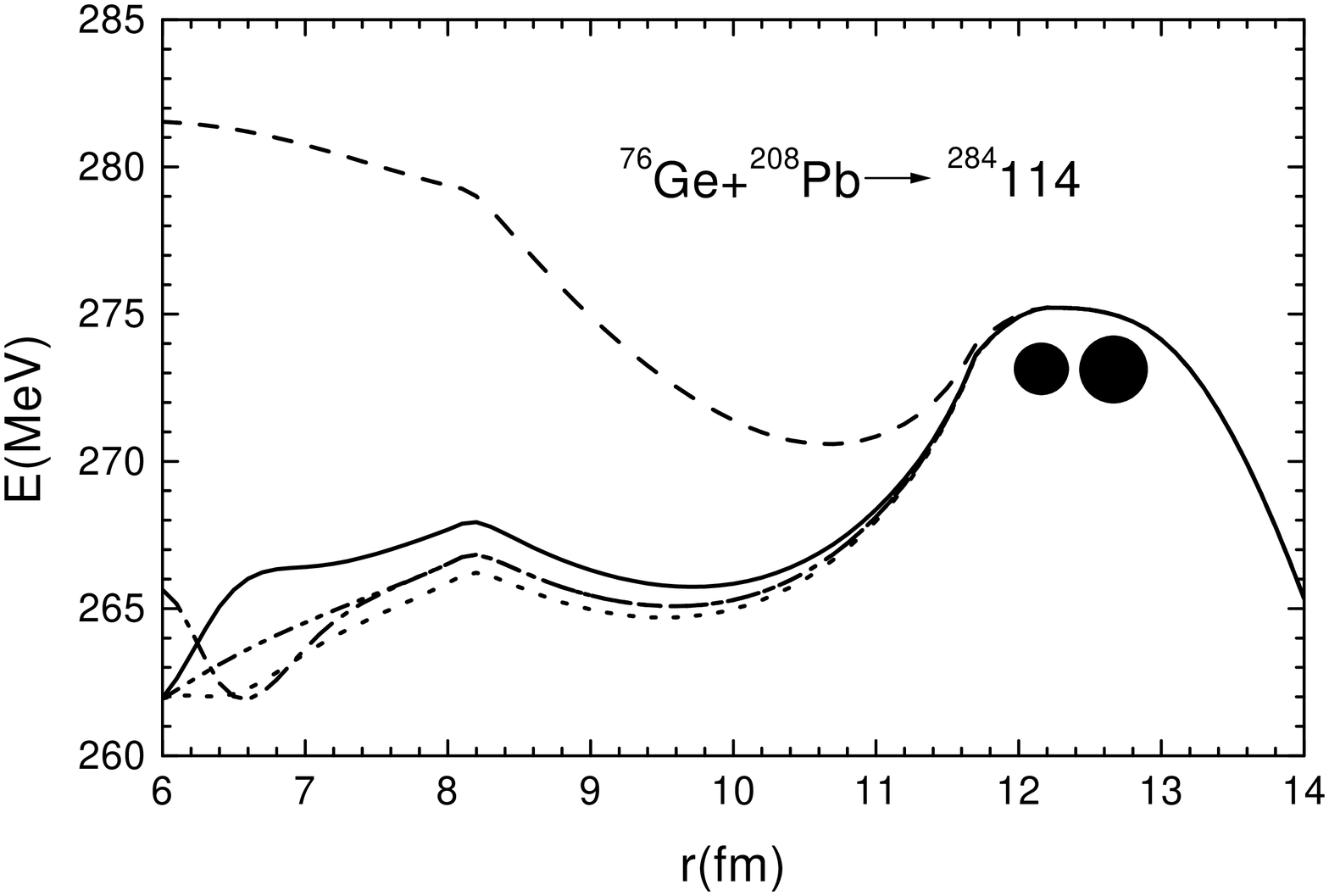}
}
\bigskip

\newpage
\centerline{
\includegraphics[width=20cm]{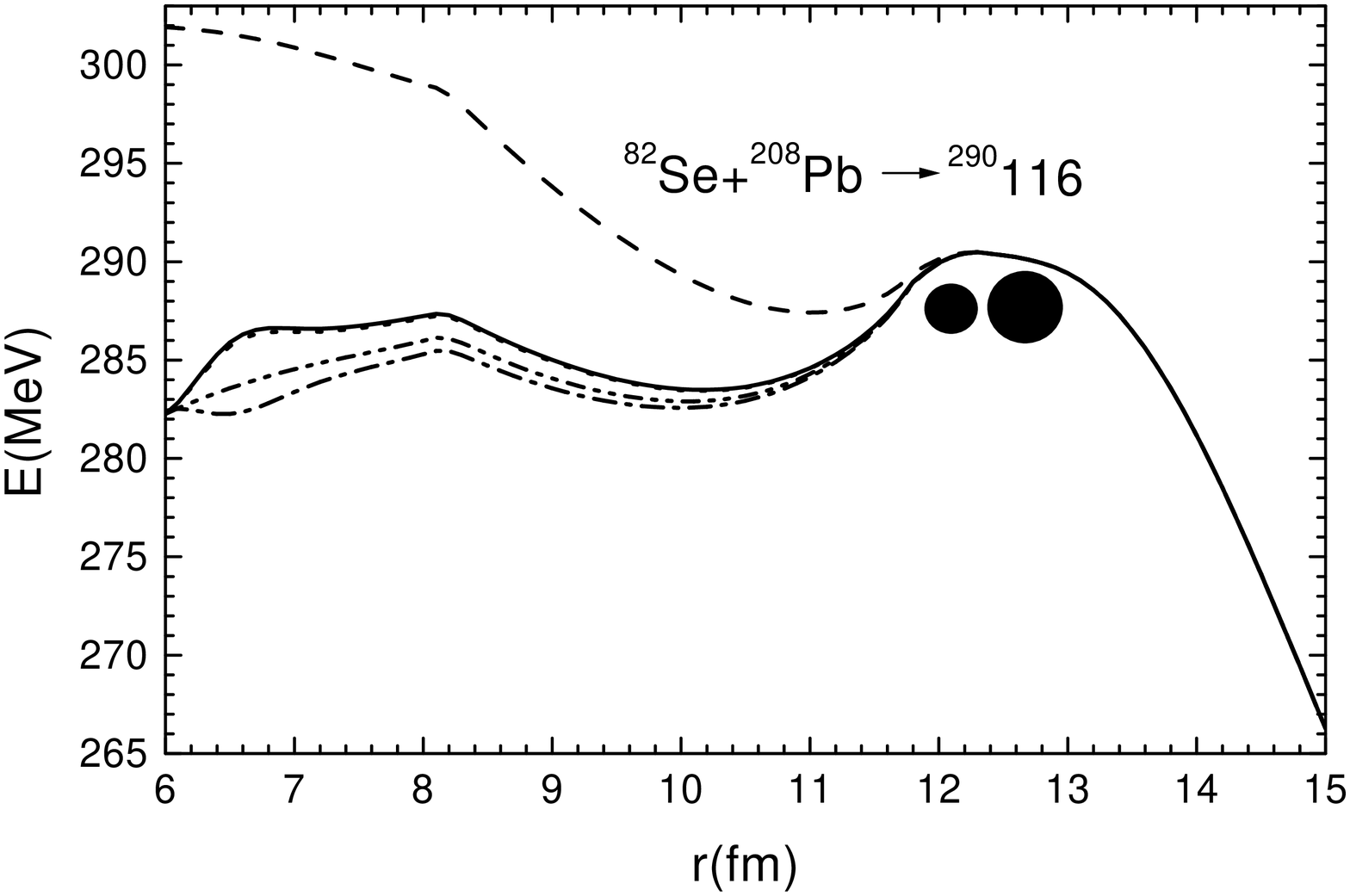}
}
\bigskip

\newpage
\centerline{
\includegraphics[width=20cm]{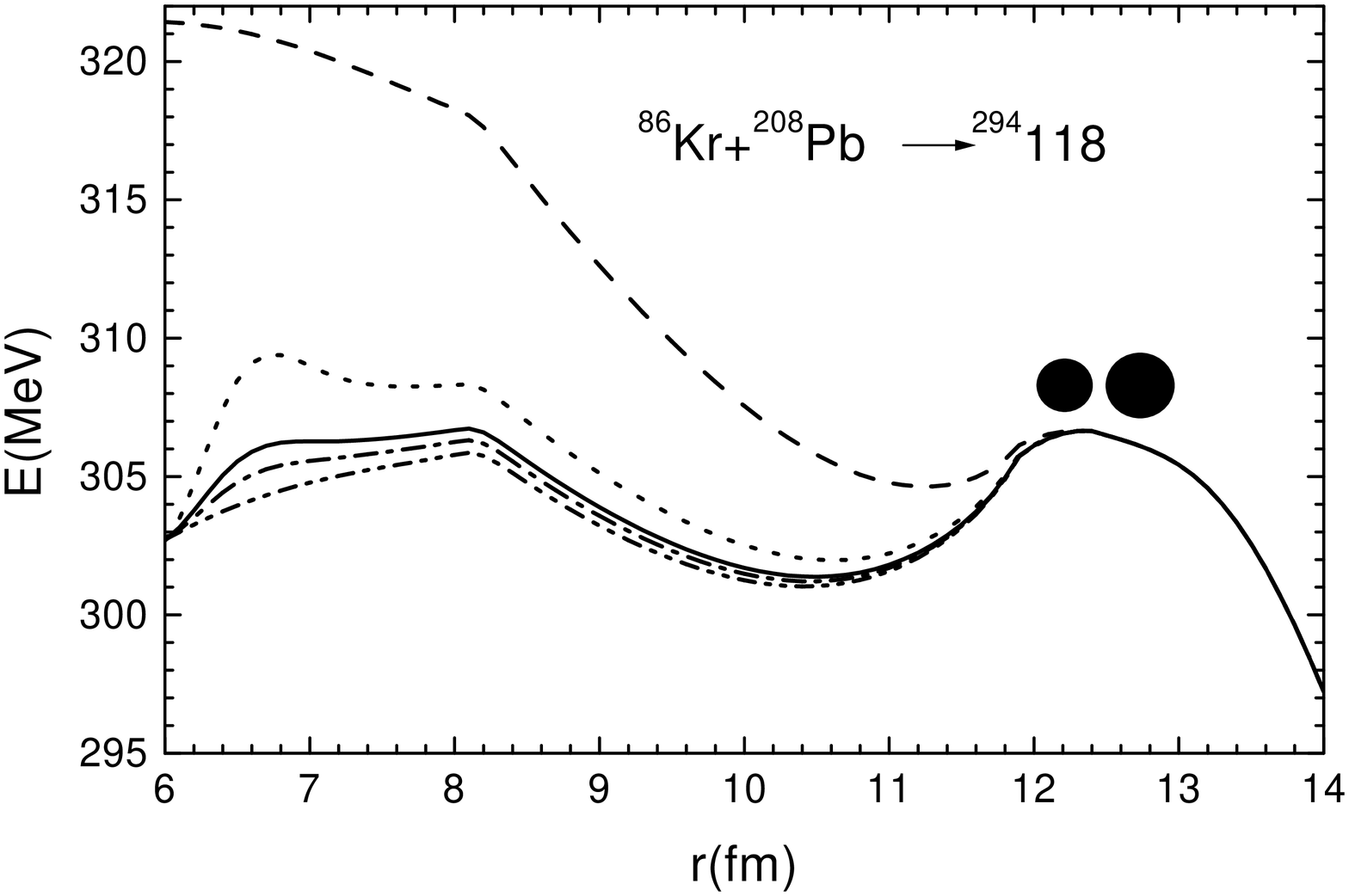}
}
\bigskip

\newpage
\centerline{
\includegraphics[width=20cm]{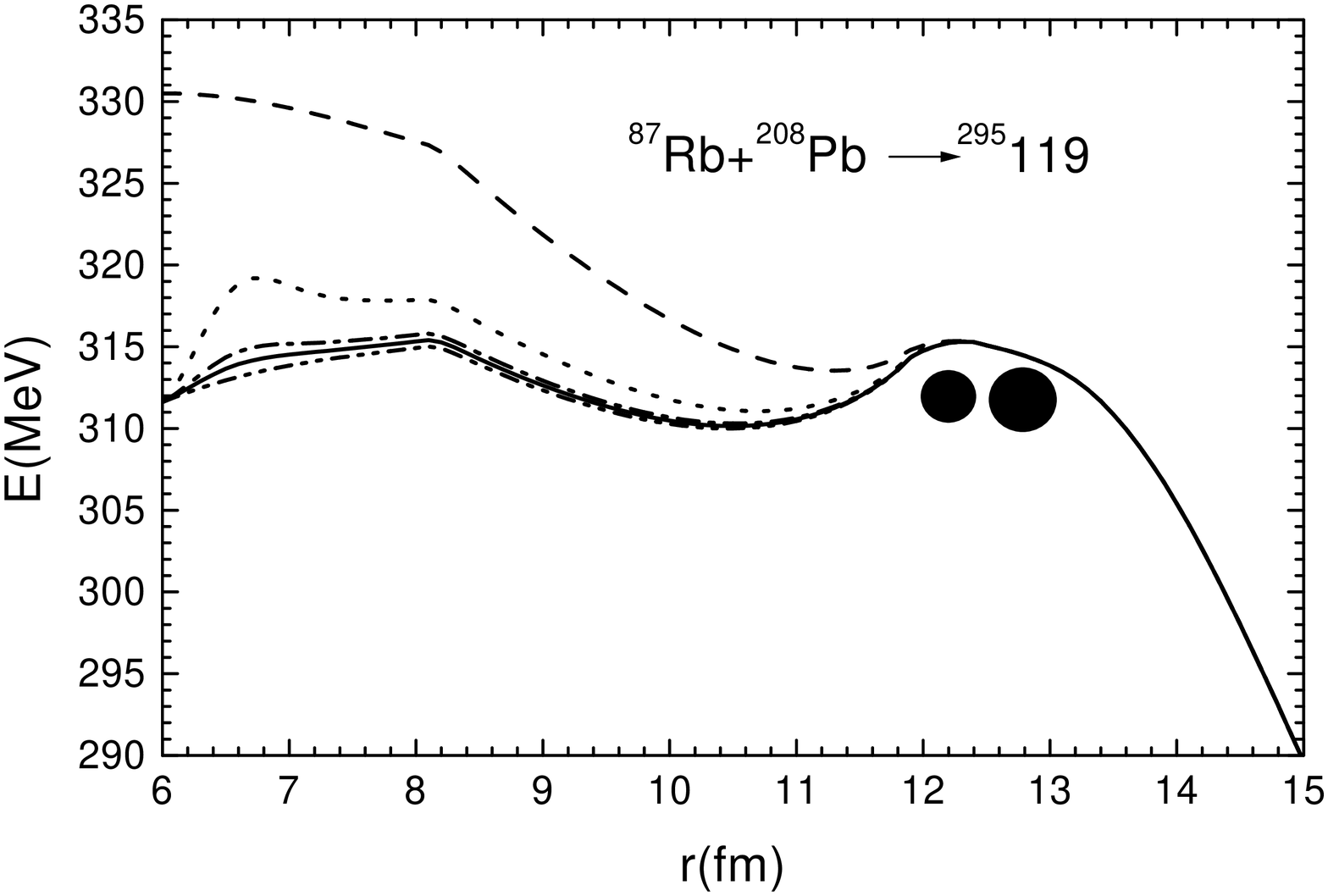}
}
\bigskip

\newpage
\centerline{
\includegraphics[width=20cm]{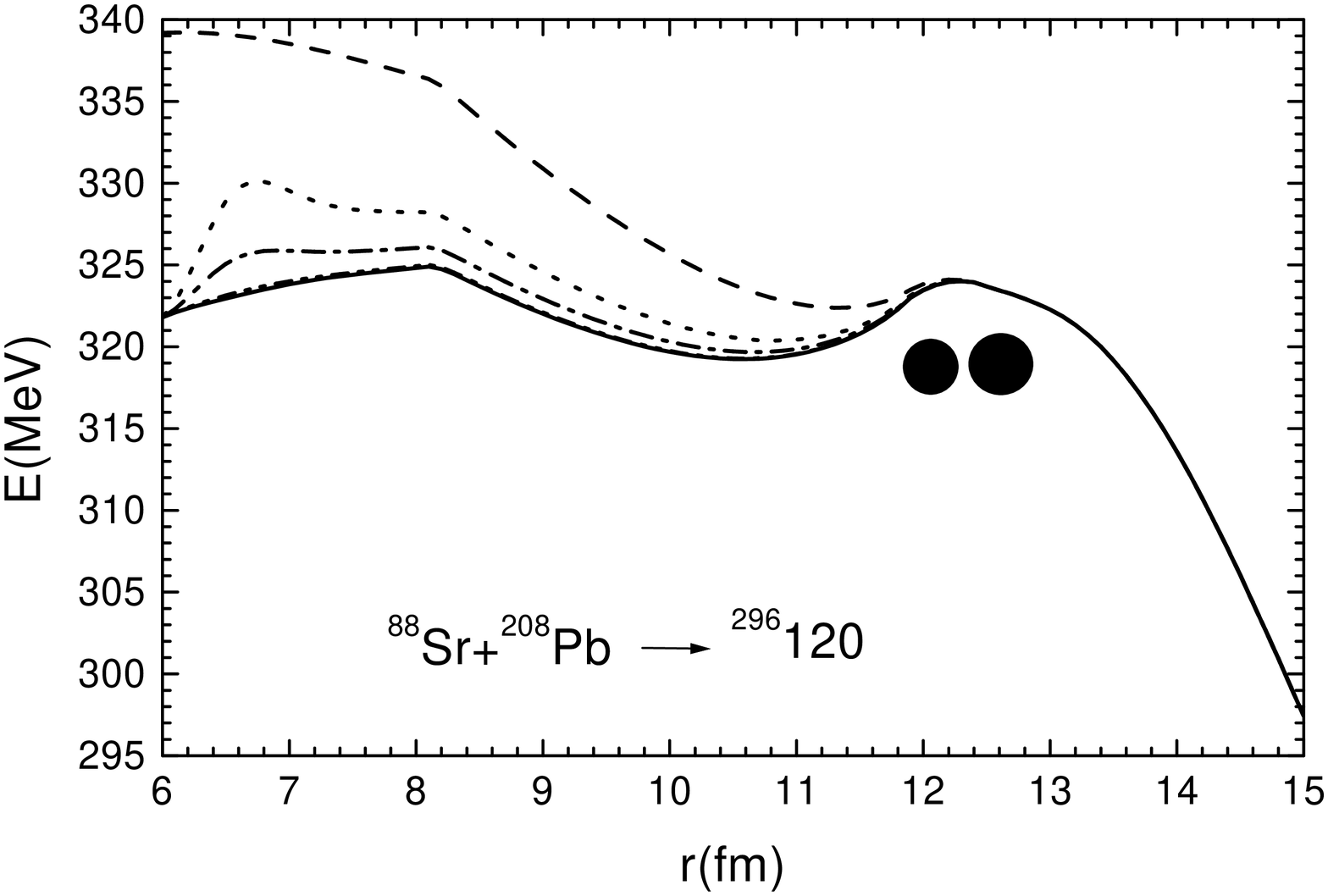}
}
\bigskip

\newpage
\centerline{
\includegraphics[width=20cm]{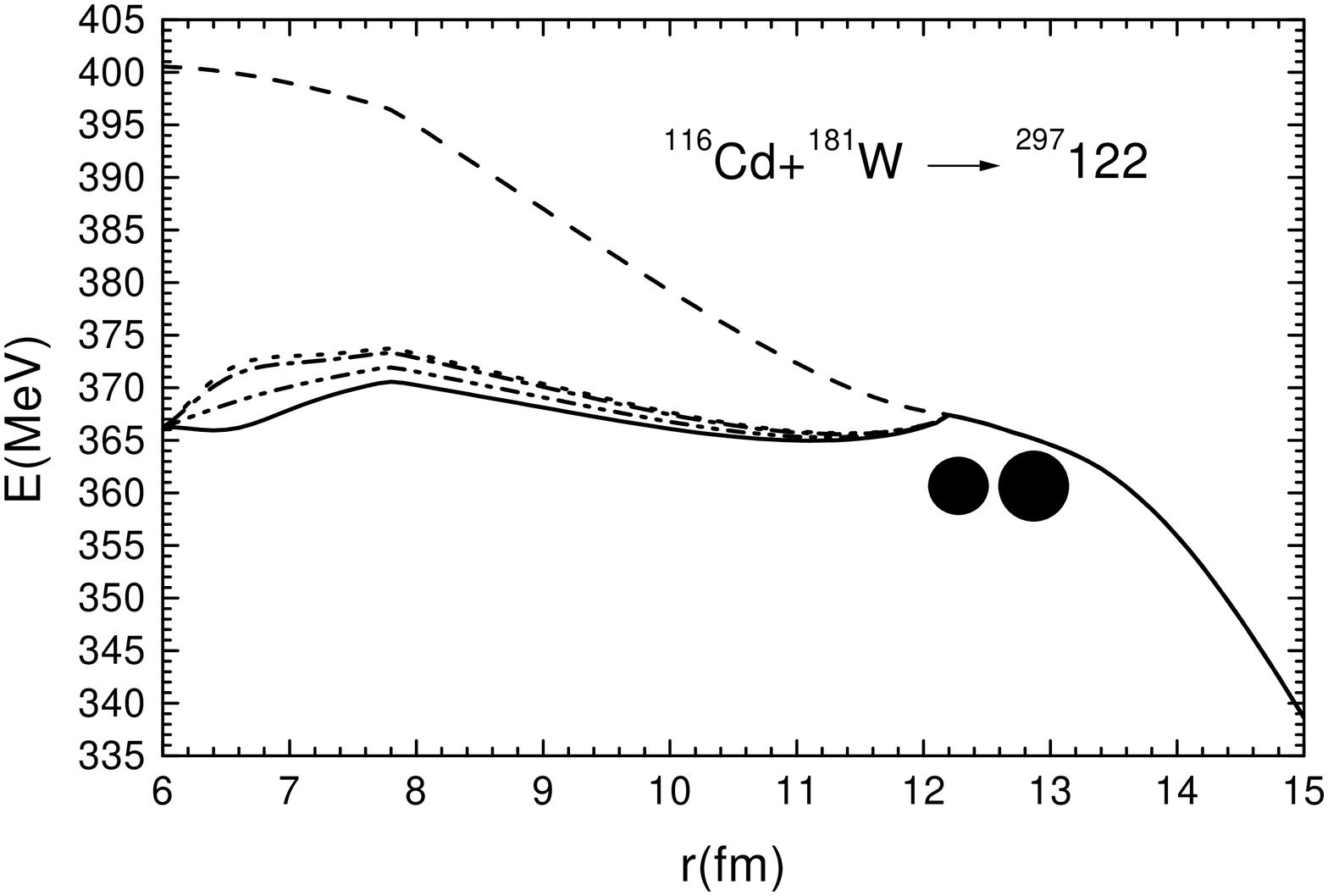}
}
\bigskip

\newpage
\centerline{
\includegraphics[width=20cm]{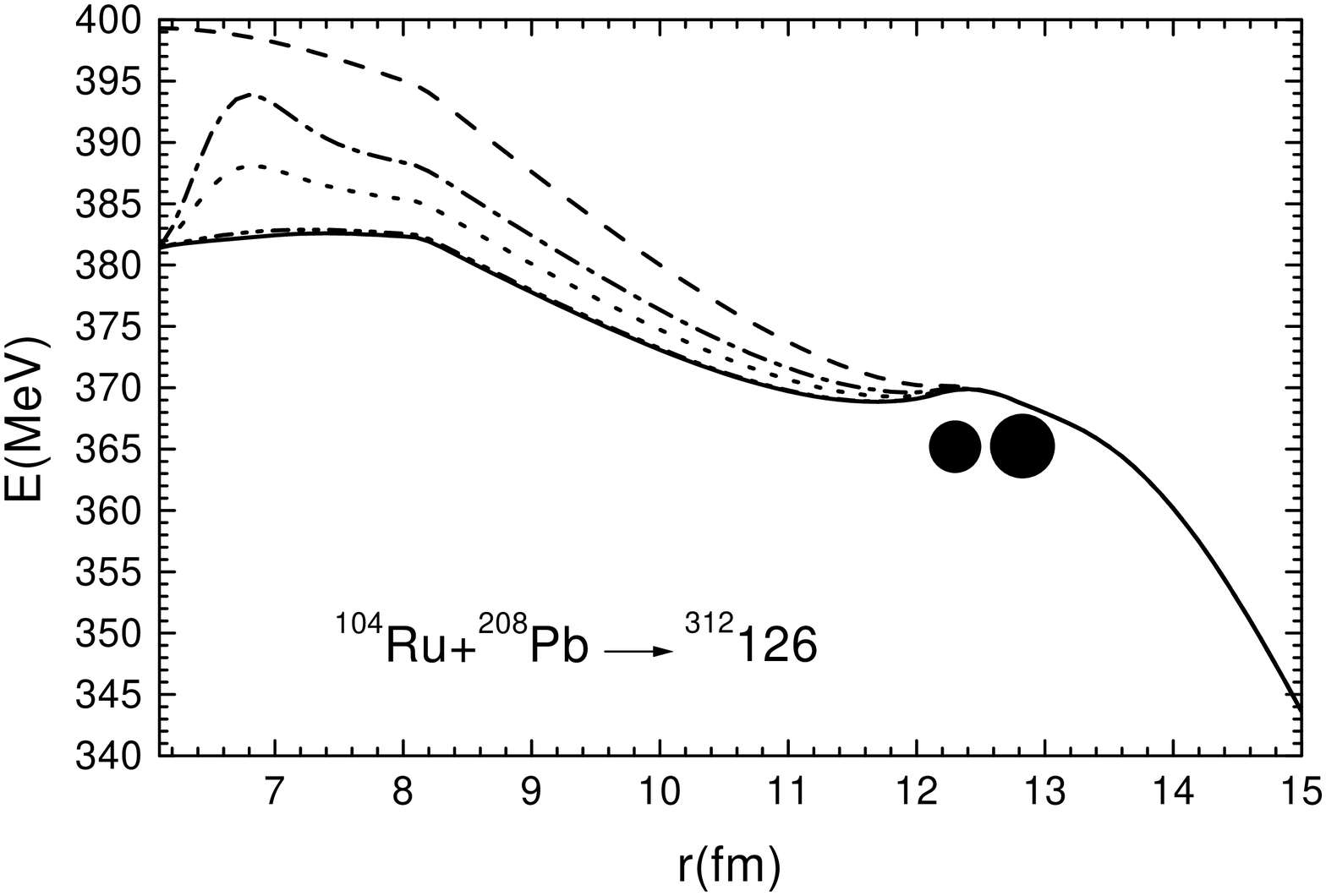}
}
\bigskip

\newpage
\centerline{
\includegraphics[width=18cm]{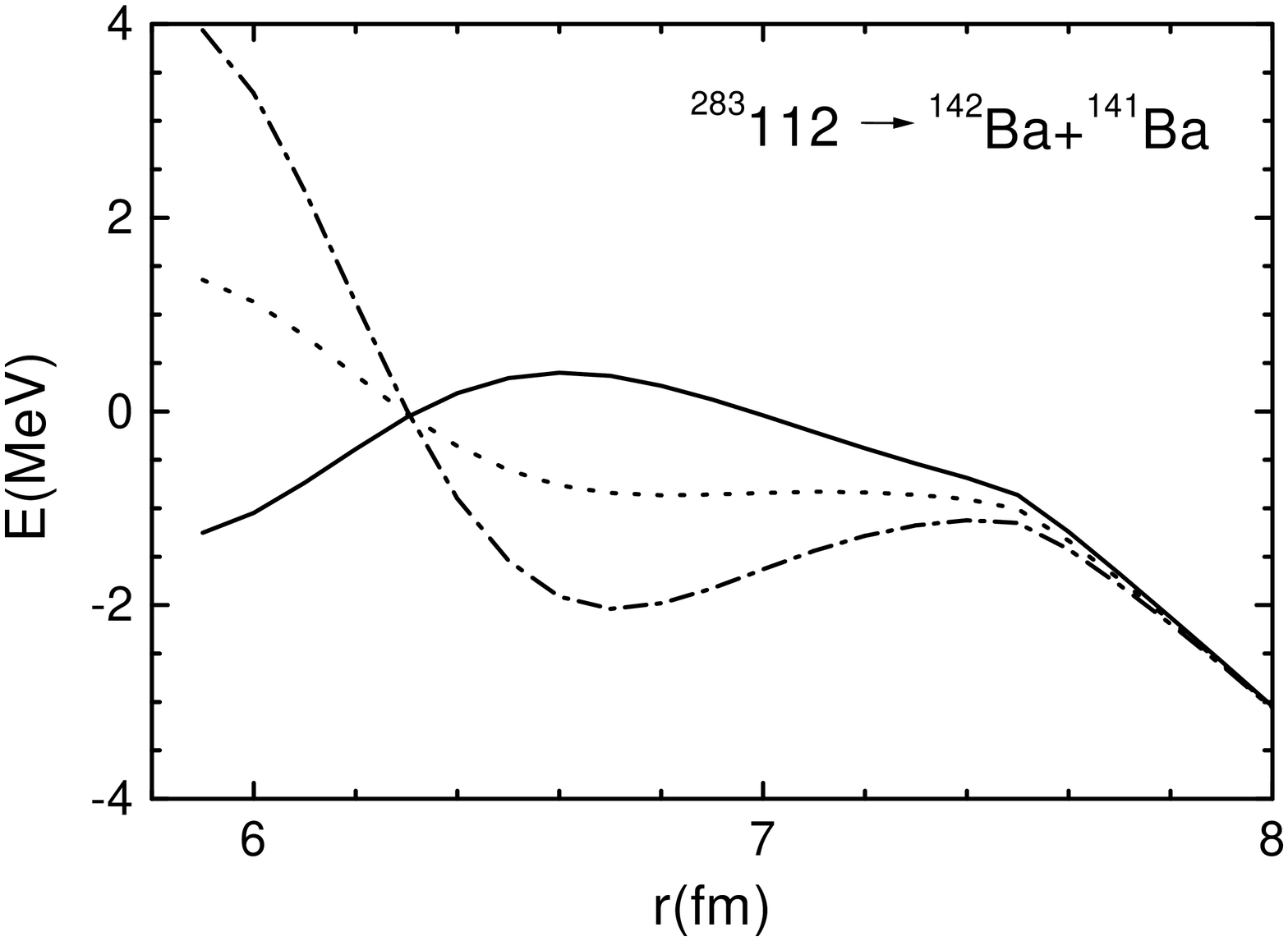}
}
\bigskip

\newpage
\centerline{
\includegraphics[width=18cm]{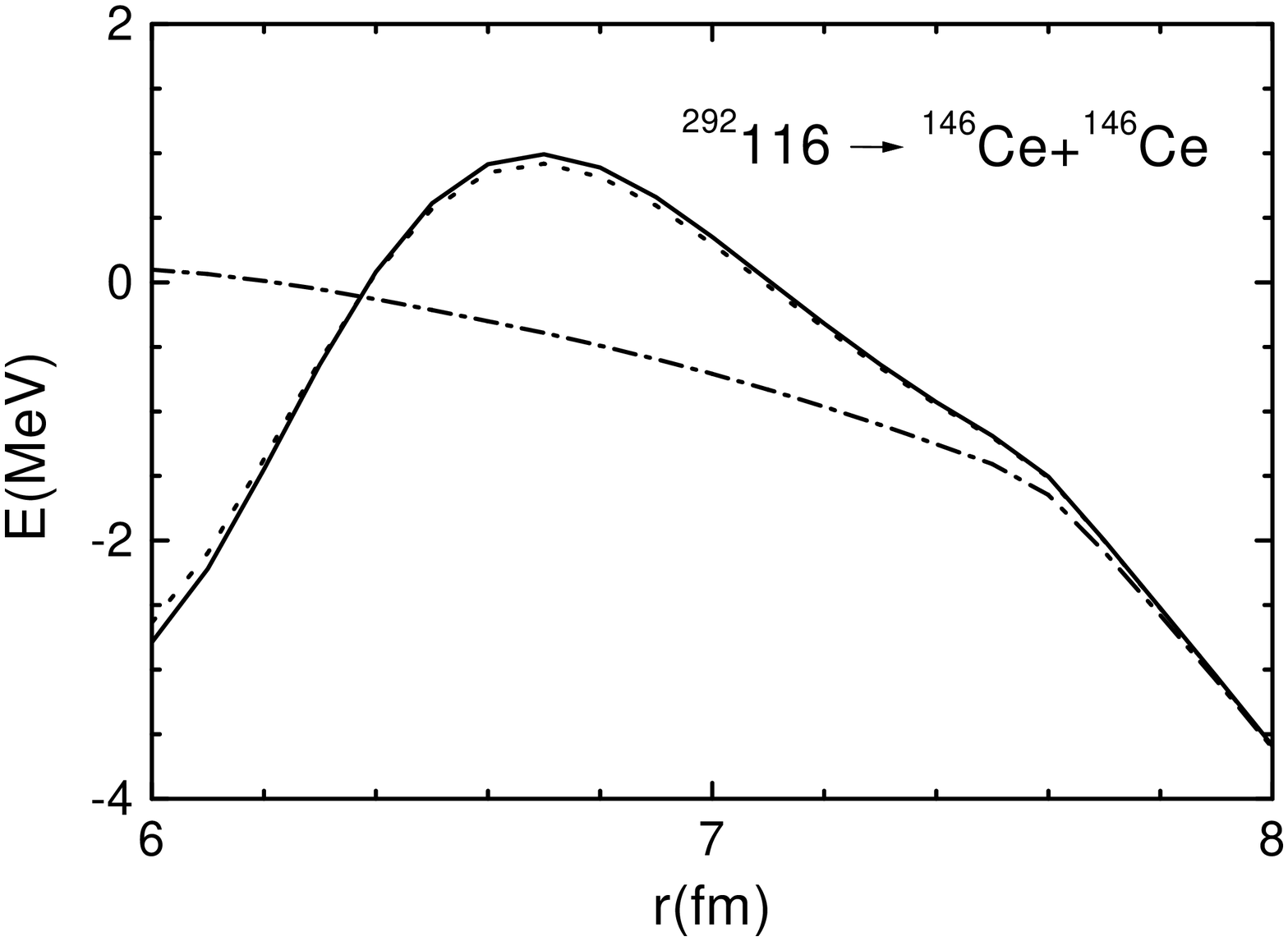}
}
\bigskip

\newpage
\centerline{
\includegraphics[width=18cm]{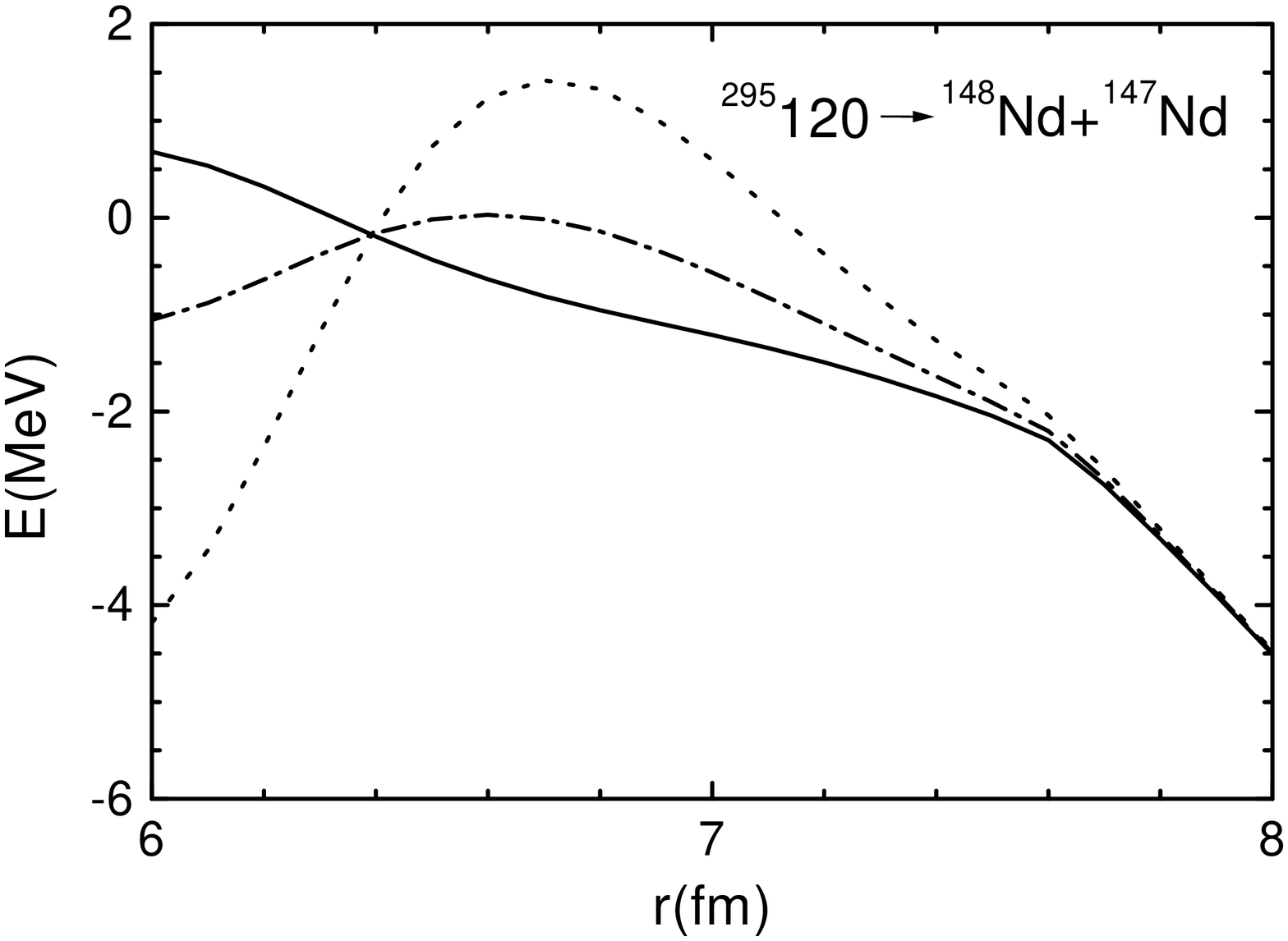}
}
\bigskip

\newpage
\centerline{
\includegraphics[width=18cm]{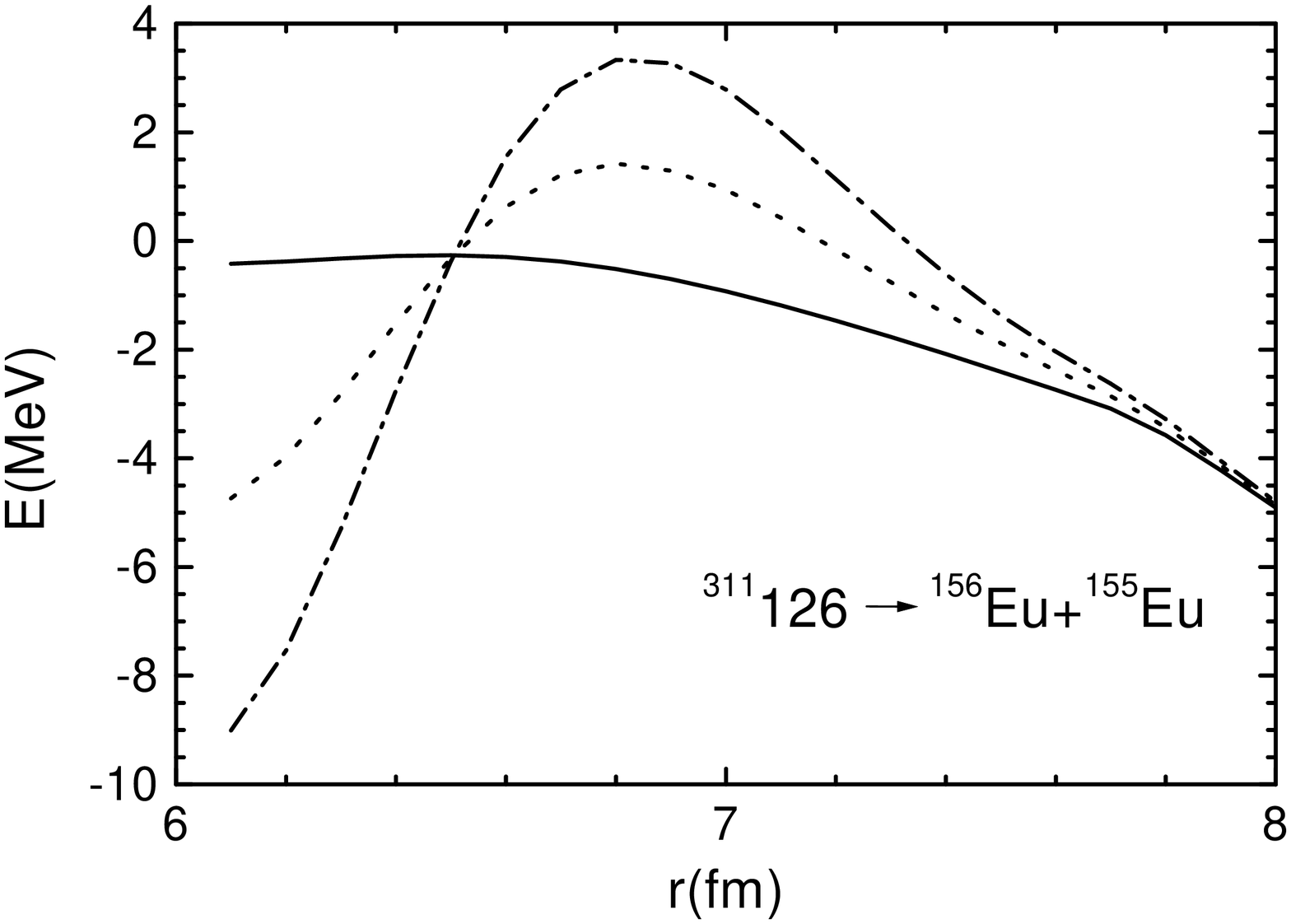}
}
\bigskip

\end{document}